\documentclass[12pt]{article}
\pdfoutput=1
\usepackage{rotating}

\usepackage{multirow}
\usepackage{amsmath,amssymb,graphicx, graphics, color,verbatim} 
\usepackage{cite}
\usepackage{latexsym}
\usepackage{cancel}

\usepackage{enumerate}

\usepackage[normalem]{ulem} 

\newcommand{\centeron}[2]{{\setbox0=\hbox{#1}\setbox1=\hbox{#2}\ifdim
\wd1>\wd0\kern.5\wd1\kern-.5\wd0\fi \copy0
\kern-.5\wd0\kern-.5\wd1\copy1\ifdim\wd0>\wd1
                                   \kern.5\wd0\kern-.5\wd1\fi}}
\newcommand{\ltap}{\>\centeron{\raise.35ex\hbox{$<$}}
                           {\lower.65ex\hbox{$\sim$}}\>}
\newcommand{\gtap}{\>\centeron{\raise.35ex\hbox{$>$}}
                           {\lower.65ex\hbox{$\sim$}}\>}
\newcommand{\gsim}{\mathrel{\gtap}}
\newcommand{\lsim}{\mathrel{\ltap}}

\newcommand\ZZ{\hbox{\zfont Z\kern-.4emZ}}
\font\zfont = cmss10 

\newcommand{\fref}[1]{Fig.\ \ref{f.#1}}
\newcommand{\eref}[1]{Eq.\ (\ref{e.#1})}

\newcommand{\sref}[1]{Section \ref{s.#1}}
\newcommand{\ssref}[1]{Section \ref{ss.#1}}

\newcommand{\cref}[1]{Chapter \ref{c.#1}}
\newcommand{\tref}[1]{Table \ref{t.#1}}

\newcommand{\ba}{\begin{array}}
\newcommand{\ea}{\end{array}}

\newcommand{\beq}{\begin{eqnarray}}
\newcommand{\eeq}{\end{eqnarray}}
\newcommand{\beqs}{\begin{eqnarray*}}
\newcommand{\eeqs}{\end{eqnarray*}}

\newcommand{\bal}{\begin{align}} 
\newcommand{\eal}{\end{align}}

\def\bi{\begin{itemize}}
\def\ei{\end{itemize}}
\def\ben{\begin{enumerate}}
\def\een{\end{enumerate}}
\def\bc{\begin{center}}
\def\ec{\end{center}}
\def\bt{\begin{table}}
\def\et{\end{table}}
\def\btb{\begin{tabular}}
\def\etb{\end{tabular}}

\def\gev{\, {\rm GeV}}

\def\tev{\, {\rm TeV}}

\def\mass2{mass${}^2$}

\def\re{{\rm Re} \,}

\begin{document}
\bibliographystyle{unsrt}
\begin{titlepage}

\vskip2.5cm
\begin{center}
\vspace*{5mm}
{\huge \bf Excluding Electroweak Baryogenesis in the MSSM}
\end{center}
\vskip0.2cm

\begin{center}
{\bf David Curtin$^{1}$, Prerit Jaiswal$^{1,2}$, Patrick Meade$^{1}$}

\end{center}
\vskip 8pt

\begin{center}
{\it $^1$ C. N. Yang Institute for Theoretical Physics\\ Stony Brook University, Stony Brook, NY 11794.}\\
\vspace*{0.3cm}
{\it $^2$ Department of Physics, Brookhaven National Laboratory\\ Upton, NY 11973, USA }\\
\vspace*{0.3cm}

\vspace*{0.1cm}

{\tt curtin@insti.physics.sunysb.edu, pjaiswal@quark.phy.bnl.gov, meade@insti.physics.sunysb.edu}
\end{center}

\vglue 0.3truecm

\begin{abstract}
In the context of the MSSM the Light Stop Scenario (LSS) is the only region of parameter space that allows for successful Electroweak Baryogenesis (EWBG).  This possibility is very phenomenologically attractive, since it allows for the direct production of light stops and could be tested at the LHC. 
The ATLAS and CMS experiments have recently supplied tantalizing hints for a Higgs boson with a mass of $\approx 125 \gev$. This Higgs mass severely restricts the parameter space of the LSS, and we discuss the specific predictions made for EWBG in the MSSM.  Combining data from all the available ATLAS and CMS  Higgs searches reveals a tension with the predictions of EWBG even at this early stage. This allows us to exclude EWBG  in the MSSM at greater than (90) 98\% confidence level in the (non-)decoupling limit, by examining correlations between different Higgs decay channels. 
We also examine the exclusion without the assumption of a $\approx 125 \gev$ Higgs. The Higgs searches are still highly constraining, excluding the entire EWBG parameter space at greater than 90\% CL except for a small window of $m_h \approx 117 - 119 \gev$.  
\end{abstract}

\end{titlepage}

\section{Introduction}
\label{s.intro} \setcounter{equation}{0} \setcounter{footnote}{0}

Baryogenesis is one of the fundamental questions left unanswered by the Standard Model (SM) of particle physics.  There are many approaches to generating the baryon asymmetry of the universe (BAU), some examples of which are electroweak baryogenesis (EWBG) \cite{firstEWBGpapers}, leptogenesis \cite{leptogenesis} and Affleck-Dine baryogenesis \cite{ADBG}.  EWBG is an intriguing possibility because it relies only upon weak scale physics and gives rise to possible direct experimental tests, but it cannot take place within the SM \cite{EWBGSM1, EWBGreviews}  given the current lower bounds on the Higgs mass~\cite{LEPhiggsbounds}.  EWBG could be realized within the Minimally Supersymmetric Standard Model (MSSM)~\cite{ SUSYEWBG}, see \cite{EWBGreviews} for reviews, but it requires a particular corner of the MSSM parameter space known as the light stop scenario (LSS)~\cite{Tc100thermalcancellation, lightstopneeded, Espinosa2loop, CQW2loop, wantthinwall, CQSWBAUcalc, ResonantRelaxation, EDMandEWBG1, CNQW2loopEFT, CNQWfullstudy, QuirosSeco2loop, Delepine:1996vn}.   As the name suggests there are in principle directly testable predictions of new light particles that can be discovered at the LHC.  However, as with many searches at the LHC, depending on the exact spectra, particles with copious production cross sections can be missed if a particular signature is not investigated.      The benefit of the LSS is that direct production of stops are not the only test of the scenario.  
In the MSSM the stop sector is crucial for a viable Higgs sector due to the needed radiative corrections to the Higgs mass.  The stops also contribute to various effective Higgs couplings, most significantly to two gluons. This intertwining of the two sectors means that there are additional tests of EWBG in the MSSM, based purely on the properties of the Higgs.

Recently, both the ATLAS \cite{ATLASgammagamma, ATLASZZ, ATLAScombination} and CMS \cite{CMSgammagamma,CMSZZ,CMScombination} experiments at the Large Hadron Collider (LHC) reported intriguing $2 - 3 \ \sigma$ excesses in the diphoton and $ZZ^* \rightarrow 4 \ell$ channels that could be interpreted as early signs of a $\approx 125 \gev$ Standard-Model-like Higgs. More data is needed to claim discovery, but it is not too early to start thinking about what the implications of such a result might be~\cite{approxLogL1, MSSMnathiggs, Carena:2011aa, Draper:2011aa, NMSSMhiggs, Cao:2012fz}.  In particular, there are only two ways that such a heavy Higgs could be realized within the MSSM: large mixing in the stop sector with relatively light stop masses, or minimal mixing with at least one of the stops being extremely heavy. 

This potential Higgs mass measurement immediately creates tension with EWBG.  A  detailed recent analysis of the LSS \cite{CNQWfullstudy} showed that  a Higgs mass of~$\approx 125 \gev$ requires the left-handed (LH) stop to be heavier than about $1000 \tev$, since the stop mixing cannot be large. The right-handed (RH) stop, on the other hand, gets pushed to a mass of $\sim 100 \gev$.  This spectrum is very peculiar, especially when trying to imagine a corresponding SUSY breaking scheme \cite{LSSinGM}. Nevertheless, it remains as the \emph{only realization of EWBG in the MSSM}.

Before the recent Higgs mass measurement, EWBG was reconcilable with a fairly wide variety of Higgs phenomenologies \cite{morrissey}. Now that the stop spectrum has been so strongly constrained by the Higgs mass measurement it may be possible to rule out EWBG purely by determining the properties of the Higgs boson. This is particularly attractive, since one could imagine many ways to hide a light stop from direct searches (e.g. decay through a displaced vertex). Using Higgs data represents a model-independent approach to excluding EWBG in the MSSM. 

We will show in this paper that the correlations between different Higgs decay channels and production modes, in particular those which occur via loops compared to those that occur at tree level, make predictions that are already in tension with the data. By combining the available constraints from LHC Higgs searches, we show that \emph{EWBG in the MSSM is already excluded at the (90) 98\% confidence level (CL)} in the (non-)decoupling limit. We also examine the exclusion without the assumption of a $\approx 125 \gev$ higgs. The higgs searches are still highly constraining, excluding the entire EWBG parameter space at the 90\% CL except for a small window of $m_h \approx 117 - 119 \gev$ due to a small excess seen in the ATLAS $\gamma\gamma$ search.

This paper is organized as follows. In \sref{review} we provide a very brief review of electroweak baryogenesis, which explains the need for the LSS in the MSSM.  We then discuss the current status of the LSS in section \sref{lss} and the particular parameter space within the LSS dictated by a Higgs of mass~$\approx 125 \gev$.  In \sref{fingerprint} we investigate the \emph{fingerprint of EWBG in the MSSM}, the correlations amongst the different production and decay channels of the Higgs that are the signature of the LSS. In \sref{expstatus} we discuss the available experimental data from LHC Higgs searches and combine them to exclude the EWBG parameter space at the 90\% CL. We conclude with \sref{conclusion}. Various technical details of the H	iggs decay width calculations are included in the Appendix.

\section{EWBG and the LSS}
\label{s.review} \setcounter{equation}{0} \setcounter{footnote}{0}

As is well known, to generate a BAU the three Sakharov conditions \cite{Sakharov} must be satisfied: $B$-violation, $CP$-violation (otherwise any $B$-producing process is cancelled out by its $CP$-counterpart) and departure from thermal equilibrium (to prevent washout of the accumulated $B$-excess). 
Electroweak Baryogenesis \cite{firstEWBGpapers} (see also \cite{EWBGreviews} for reviews)   is a mechanism for producing the BAU that relies entirely on weak scale physics to satisfy the Sakharov conditions.  The triangle anomaly in the electroweak sector of the Standard Model leads to non-perturbative sphaleron processes at high temperatures that violate baryon number, complex phases in the Higgs-fermion couplings provide the necessary $CP$-violation, and departure from thermal equilibrium is instigated by the electroweak phase transition.  Even though the Standard Model has all the qualitative ingredients for electroweak baryogenesis,  the size of the generated baryon asymmetry falls far short of the required value \cite{EWBGSM1, EWBGreviews}. The phase transition is first order but only weakly so, and there is not enough $CP$ violation. This means that electroweak baryogenesis can only work in a theory with additional complex phases, as well as weak-scale particles that interact strongly with the Higgs sector to give the necessary additional contributions to its thermal potential. These conditions can be satisfied in the MSSM \cite{SUSYEWBG}: contributions from stops to the thermal potential of the Higgs can generate a stronger first order phase transition, and there are many new sources of CP violation available.  

The task of computing the generated BAU can be approximately factorized, into
sectors that are responsible for the first order phase transition, and those directly responsible for creating the baryon asymmetry during that phase transition. Therefore we can examine the constraints or evidence for these sectors independently.  

Provided that a strong enough first order phase transition occurs, 
computation of the BAU involves a complicated tunneling, quantum transport and hydrodynamics calculation. $CP$-violating interactions between the plasma and the space-time varying Higgs VEV generate chiral currents across the bubble wall,  which sphalerons in the unbroken phase convert to baryon asymmetry. This excess then partially survives in the broken phase, where sphalerons are suppressed, as the bubble expands. There is a vast literature on this calculation~\cite{
wantthinwall, bubblecalc1, FirstResonantQTE, ResonantRelaxation, GammaYbreakdown, superequillibriumOK, CQSWBAUcalc, BinoDrivenEWBG, Cline:1997vk, Cline:2000nw
}. The uncertainties  are still order one, and tend to err on the optimistic side \cite{Yboverestimated}. That being said, the generation of a sufficient BAU seems at least possible within the MSSM for some suitably chosen gaugino/higgsino parameters ($M_1, M_2, \mu, \tan \beta, m_A$),  \emph{if} there is a strong enough first order phase transition.\footnote{
It was recently suggested~\cite{Barenboim} that modifying the thermal history of the universe could enlarge the parameter space for EWBG within the MSSM. However, given the known mechanisms for  generating baryons during the phase transition, this is not a viable proposal. 
}

The only severe constraint from this step of the calculation comes from EDMs \cite{EDMreview} that can arise as a result of the required  $CP$-violating phases. The required phases are $\phi_1 = \mathrm{Arg}(\mu M_1 b^\star)$ and $\phi_2 = \mathrm{Arg}(\mu M_2 b^\star)$, where $b$ is the Higgs sector soft mass. One-loop EDM contributions can be suppressed by making the first and second generation sfermions heavier than $\sim 10 \tev$, but two-loop contributions involving the chargino and Higgs fields are sizable unless $m_A \gsim 1 \tev$ (see e.g. \cite{EDMandEWBG1, CNQWfullstudy, BinoDrivenEWBG}).  This generic bound can be loosened if $\phi_2$ is strongly suppressed relative to $\phi_1$, since the phase of the bino-mass by itself does not generate strong two-loop EDM contributions. In this \emph{bino-driven} scenario \cite{BinoDrivenEWBG} $m_A$ can take on smaller values.

Calculating the strength of the first order phase transition is somewhat more straightforward, and ultimately more constraining, than the baryon density calculation.  A sufficiently strong phase transition requires  $v_c/T_c \gsim 1$ (see e.g. \cite{EWBGreviews}),  where $T_c \approx 100 \gev$ is the \emph{critical temperature} at which the electroweak symmetry breaking vacuum $\phi = v_c$ is degenerate with the symmetric minimum $\phi = 0$. In the one-loop thermal Higgs potential  one finds that $v_c/T_c \sim$~(cubic coefficient)/(quartic coefficient). The cubic term comes solely from the thermal contribution and has the form $\delta V\sim T m_i(\phi)^3$, where $m_i(\phi)$ is the field dependent thermal mass of the additional scalars in the MSSM.  To maximize the strength of the phase transition clearly requires maximizing the new contributions to the cubic term.  Given the form of the contribution from the scalars of the MSSM, the largest potential contribution will come from the stop sector.  The Higgs dependent masses of the stops are given by
\begin{eqnarray*}
m_{\tilde t_R}^2 &=& m_{Q_3}^2 + h_t^2 \phi_u^2 + \left(\frac{1}{2}  - \frac{2}{3} \sin^2\theta_W\right) \frac{g^2+{g^\prime}^2}{2} (\phi_u^2 - \phi_d^2)\\
m_{\tilde t_L}^2 &=& m_{U_3}^2 + h_t^2 \phi_u^2 + \left( \frac{2}{3} \sin^2\theta_W\right) \frac{g^2+{g^\prime}^2}{2} (\phi_u^2 - \phi_d^2)\\
m_X^2 &=& h_t (A_t \phi_u - \mu \phi_d)
\end{eqnarray*}
where $\phi_{u, d} = \re H_{u,d}^0$. Working along the direction of the zero-temperature Higgs-VEV in the Higgs potential\footnote{This is valid at the critical temperature if $m_A$ is large, and sufficient for our purposes of demonstrating the effect of stops on $V_{eff}(\phi_u, \phi_d)$.}, $(\phi_u, \phi_d) = (\phi \sin \beta , \phi \cos \beta )$, their mass eigenvalues are
\begin{equation}
m_{\tilde t_{1,2}}^2(\phi) = \frac{m_{\tilde t_L}^2(\phi) + m_{\tilde t_R}^2(\phi)}{2} \pm \sqrt{
\left(\frac{m_{\tilde t_L}^2(\phi) - m_{\tilde t_R}^2(\phi)}{2}\right)^2 + \left[ m_X^2 (\phi)\right]^2. 
}
\end{equation}
In the cubic term of the thermal one-loop Higgs potential one has to replace the soft masses $m^2_{Q_3, U_3}$ by $m^2_{Q_3, U_3} + \Pi_{t_L, t_R}$, where $\Pi_{t_L}$, $\Pi_{\tilde t_R} \sim g^2 T^2$ are the thermal masses of the LH and RH stops. (This is necessary to  control IR divergences in the one-loop thermal potential and restore the validity of the perturbative expansion at the critical temperature \cite{QuirosDaisy, ThermalMasses}).  To maximize the cubic Higgs term, one of the stop mass eigenvalues should therefore be close to $\sim h_t^2 \phi^2$. This requires small stop mixing, as well as a fairly precise cancellation between the light stop's thermal mass, $\Pi_{\tilde t_R} \sim g^2 T_c^2$, and the necessarily negative stop soft mass-squared \cite{Tc100thermalcancellation}. This yields one stop that is lighter than the top. To increase the Higgs mass beyond the LEP limit \cite{LEPhiggsbounds} and avoid large corrections to the $\rho$-parameter \cite{rhoparam} the LH stop should then be heavier than a TeV or so.

Two-loop corrections \cite{Espinosa2loop, CQW2loop, CNQW2loopEFT, QuirosSeco2loop, Cline:1998hy} are quite large because $g_s$ enters for the first time at this order. Both two-loop and non-perturbative corrections \cite{lightstopneeded, Cline:1996cr} enhance the  phase transition, enlarging the viable parameter space. This provides a more complete picture of the viable regions of MSSM parameter space for electroweak baryogenesis, but the intuition from examining one-loop effects still provides a helpful guide.

Putting all these ingredients together leads to the \emph{Light Stop Scenario} (LSS)\cite{Tc100thermalcancellation, lightstopneeded, Espinosa2loop, CQW2loop, wantthinwall, CQSWBAUcalc, ResonantRelaxation, EDMandEWBG1, CNQW2loopEFT, CNQWfullstudy, QuirosSeco2loop, Delepine:1996vn}, the only corner of MSSM parameter space where electroweak baryogenesis might be possible. The constraints on the stop sector parameters  are the following:
\begin{itemize}
\item Achieving a strong phase transition and avoiding color-breaking requires a mostly right-handed light stop with $m_{\tilde t_1} < m_t$ and $A_t \lsim m_Q/2$. 
\cite{Tc100thermalcancellation, CNQW2loopEFT, CNQWfullstudy}.
\item The mostly left-handed stop should be heavier than $\sim \tev$ to satisfy the LEP Higgs mass bound (for a SM-like Higgs) and avoid large corrections to the $\rho$-parameter. 
\item The gluino should be heavier than $\sim 500 \gev$ to decouple it from the plasma, otherwise its large contribution to the stop thermal masses would make it even more difficult to achieve the needed cancellation $m_{U_3}^2 \sim - \Pi_{t_R}$.
\end{itemize}
In addition, there are some constraints on the electroweak gaugino and higgsino parameters to allow for sufficient generation of BAU: 
\begin{itemize}
\item  $M_1$ or $M_2 \sim \mu \sim \mathcal{O}(100 \gev)$ with sufficiently large $CP$-violating phases in the -ino sector, as well as $\tan \beta \lsim 15$ \cite{CNQWfullstudy}. 
\item $m_A \gsim 1 \tev$, unless all the $CP$-violation is pushed into the bino soft mass \cite{BinoDrivenEWBG}.
\end{itemize}

\section{LSS and a heavy Higgs}
\label{s.lss} \setcounter{equation}{0} \setcounter{footnote}{0}

It is well known to experts that the possible detection of a $\approx 125$ GeV Higgs at the LHC  spells trouble for electroweak baryogenesis in the MSSM.  Such a large Higgs mass can only be achieved if stop mixing is large (incompatible with a strong phase transition) or if the left-handed stop is extraordinarily heavy. To quantify the exact consequences of such a heavy Higgs, we draw upon the results of \cite{CNQWfullstudy}.

The authors of \cite{CNQWfullstudy} studied the electroweak baryogenesis window of the MSSM in great detail. They constructed a low-energy effective theory \cite{CNQW2loopEFT} in which all scalar superpartners with the exception of the RH stop are pushed to some high common scale.\footnote{
The scalars other than the LH stop have been made heavy to satisfy EDM constraints, but this might not be necessary (e.g.  Bino-Driven EWBG \cite{BinoDrivenEWBG}). Nevertheless, the derived restrictions on the stop spectrum should be widely applicable.}
This effective description, tailored to the Light Stop Scenario, included the most important one- and two-loop effects. They constructed the resulting thermal Higgs potential and scanned over the stop- and Higgs-sector parameter space. Requiring a sufficiently strong first-order phase transition and avoiding color-breaking yields regions of the stop-Higgs mass plane where electroweak baryogenesis could proceed within the MSSM. 

As expected, a Higgs mass in the range of $123 \gev \leq m_h \leq128 \gev$ is extremely difficult to accommodate. The stop sector has to take on a very particular form:
\begin{equation}
\label{e.stopspectrum}
m_{\tilde t_R} = 80 - 115 \gev \ \ , \ \ \ \ \ m_{\tilde{t}_L} \gsim 10^3 \tev \ \ , \ \ \ \ \ \tan \beta \approx 5 - 15
\end{equation}
with stop mixing being completely negligible for such large $m_Q$. The size of the allowed $m_{\tilde t_R}$ range is somewhat overestimated, since it was obtained by interpreting the results of the analysis in a \emph{very} conservative fashion. Therefore, if this stop spectrum can be excluded, then electroweak baryogenesis in the MSSM is excluded (assuming of course that the Higgs mass falls into the above mentioned range).

The extremely heavy left-handed stop is in significant conflict with notions of naturalness, reminiscent  of Split Supersymmetry \cite{splitsusy}. One could ask how a high-energy theory of SUSY breaking could generate such a spectrum \cite{LSSinGM}, but let us put aside such considerations and focus on the phenomenology. 

The light right-handed stop with a mass of $\sim 100  \gev$ is an extremely interesting \emph{prediction} of electroweak baryogenesis within the MSSM, emerging as a direct consequence of requiring a sufficiently strong electroweak phase transition and a Higgs mass of $\approx 125 \gev$. It is already excluded if it decays promptly \cite{promptstopsearchD0, promptstopsearchCDF, promptstopNLSPKS} or escapes the detector \cite{champconstraints}, but one could imagine it being hidden from direct stop searches somehow, for example by decaying via a displaced vertex \cite{ATLASdisplacedvertexsearch}.

The question is then: given a Higgs mass of $\approx 125 \gev$, can electroweak baryogenesis within the MSSM be excluded in a model-independent way? As it turns out, the answer is yes. The specific spectrum required by the LSS in light of such a Higgs mass, especially the light RH stop, makes very definite predictions for the Higgs production rate and branching ratios. This allows us to test electroweak baryogenesis within the MSSM using pure Higgs  data, separate from collider searches for the stop and questions of how such a strange spectrum could be generated by a high-energy theory of SUSY-breaking.

\section{The Fingerprint of Electroweak Baryogenesis}
\label{s.fingerprint} \setcounter{equation}{0} \setcounter{footnote}{0}

The presence of a light RH stop can significantly alter Higgs production and decay rates compared to their SM expectation. In the context of overall inclusive production cross sections this has been investigated in detail by~\cite{morrissey}. However, even without an unambiguous $5\sigma$ Higgs discovery at the LHC, or an extremely  precise measurement of the $\gamma\gamma$ branching fraction, it is {\em still} possible to conclusively test the mechanism of EWBG in the MSSM.  This is because the LSS makes specific predictions for \emph{all} possible production and decay modes of the Higgs, and they have very particular correlations.

The presence of the light RH stop affects Higgs phenomenology through loop level production via gluon fusion and decays to $\gamma\gamma$.  The effects are encoded by examining the partial widths, which can be related to both production and decay. The leading order contributions to gluon fusion (in the decoupling limit) are \cite{Djouadi:2005gj}
\begin{equation}
\Gamma (h \rightarrow g g) = \frac{G_{\mu} \alpha_s^2 m_h^3}{36 \sqrt{2} \pi^3} \left| \frac{3}{4} \sum\limits_{f} A_{1/2} (\tau_f) +  \frac{3}{4} \frac{g_{h \tilde{t}_R \tilde{t}_R}}{m_{\tilde{t}_R}^2 }A_0 (\tau_{\tilde{t}_R}) \right|^2,
\label{e.hggdecay}
\end{equation}
where $\tau_i = m_h^2/4 m_i^2$, $g_{h \tilde{t}_R \tilde{t}_R}$ is the normalized Higgs coupling to the right-handed stop, which in the LSS is given by
$g_{h \tilde{t}_R \tilde{t}_R} \approx m_t^2 + 2/3 \cos{2\beta} s_w^2 m_Z^2$.  The functions $A_s$ ($s = 0, 1/2$ or $1$) are defined as
\begin{eqnarray}
A_0(\tau)&=& -\left[\tau-f(\tau)\right]/\tau^2\\
A_{1/2}(\tau)&=& 2 \left[\tau+(2\tau-1)f(\tau)\right]/\tau^2\nonumber\\
A_1(\tau)&=& -\left[2\tau^2+3\tau+3(2\tau-1)f(\tau)\right]/\tau^2\nonumber
\end{eqnarray}
where
\begin{equation}
f(\tau)=\left\{ \begin{array}{cc}
\arcsin^2\sqrt{\tau} & \tau\leq 1\\
-\frac{1}{4}\left[\log\frac{1+\sqrt{1-\tau^{-1}}}{1-\sqrt{1-\tau^{-1}}}-i\pi\right]^2 & \tau > 1
\end{array}\right. .
\end{equation}
The crucial point is that the light stop loop interferes constructively with the top quark loop, which leads to a more than three-fold {\em increase} in the Higgs production cross section via gluon fusion.  However, when investigating the clean $\gamma\gamma$ decay channel we must also examine the stop's contribution to the $h \rightarrow \gamma \gamma$ decay width, which at lowest order (again in the decoupling limit) is
\begin{equation}
\begin{split}
\Gamma (h \rightarrow \gamma \gamma) &= \frac{G_{\mu} \alpha^2 m_h^3}{128 \sqrt{2} \pi^3} \biggl|  \sum\limits_{f} N_c Q_f^2 A_{1/2} (\tau_f) + A_1(\tau_W) \biggr. \\
 &\quad \biggl. + \frac{4}{3} \frac{g_{h \tilde{t}_R \tilde{t}_R}^2}{m_{\tilde{t}_R}^2 }A_0 (\tau_{\tilde{t}_R}) + \sum\limits_{\chi^+} \frac{2m_W}{m_{\chi^+}} g_{h \chi^+ \chi^-} A_{1/2}(\tau_{\chi^+}) \biggr|^2
\label{e.hgagadecay}
\end{split}
\end{equation}
where we have also included the contribution from Charginos.  We do not explicitly calculate $\Gamma (h \rightarrow \gamma \gamma)$ as a function of the chargino mass, but they can shift the width at most by order $10\%$ if their mass is small, $m_{\chi^+} \sim 100$ GeV, and we include this as a theory uncertainty.  Unlike for the gluon width, the stops destructively interfere with the dominant contribution in \eref{hgagadecay} coming from the $W$ bosons, and thus {\em  decrease} the decay width $\Gamma (h \rightarrow \gamma \gamma)$ by nearly a factor of $1/2$ compared to the SM expectation. 

Thus, while a light stop can effect both $\Gamma (h \rightarrow g g)$ and $\Gamma (h \rightarrow \gamma \gamma)$ significantly, the effect can be washed out by looking at the total rate of $\sigma (gg\rightarrow h \rightarrow \gamma \gamma)$ only.  However, there are both additional Higgs production and decay modes available at the LHC.  

In particular, examining both tree level and loop level (affected by the stop) production and decay modes in various combinations should reveal strong correlations amongst the various channels. We call this the \emph{fingerprint of electroweak baryogenesis in the MSSM}.
For example, in gluon fusion events with tree level decays one would expect a large increase over Standard Model rates.  However, if one examined vector boson fusion production of a Higgs which then decayed to $\gamma\gamma$, we would expect a rate smaller than the SM.  
For EWBG we can give  predictions for all the channels, and even without a Higgs discovery one can still make strong exclusion statements by considering their correlations.   This is similar in spirit to analyses investigating naturalness and general composite higgs sectors~\cite{approxLogL1,approxLogL2}.

We use \cite{Djouadi:2005gi, Djouadi:2005gj} to compute the EWBG predictions for the Higgs decay widths. To understand the various correlations and predictions for EWBG quantitatively we define ratios of the various production channels, gluon fusion (ggF), vector boson fusion (VBF) and associated production (AP), in the LSS compared to the SM:
\begin{equation}\label{e.crosssectionratios}
r_{ggF} \equiv \frac{\sigma_{MSSM}(gg \rightarrow h) }{\sigma_{SM}(gg \rightarrow h)},\quad
r_{VBF} \equiv \frac{\sigma_{MSSM}(VBF) }{\sigma_{SM}(VBF)}, \quad r_{AP} \equiv \frac{\sigma_{MSSM}(AP) }{\sigma_{SM}(AP)}.
\end{equation}
$r_{ggF}$ is derived by taking ratios of decay widths, while $r_{VBF}, r_{AP} \approx 1$ in the decoupling limit. Similarly we can define ratios for the branching fractions $h \rightarrow X$ compared to the SM as:
\begin{equation}
b_X = \frac{ \mathrm{Br}_{MSSM}(h \rightarrow X)}{ \mathrm{Br}_{SM}(h \rightarrow X)}.
\end{equation}
Combining these various production and decay channels we can define a \emph{partial signal strength}
\begin{equation}
\label{e.partialsigstrength}
\mu_{X(i)} = \frac{\sigma(i\rightarrow h\rightarrow X)}{\sigma(i\rightarrow h\rightarrow X)_{SM}},
\end{equation}
where $X$ labels the Higgs decay final state and $i$ represents the production channel.  
As an example, our partial signal strengths for $\gamma\gamma$ final states when produced through gluon fusion and vector boson fusion are
\begin{equation}
\label{e.partialsigstrength2}
\mu_{\gamma \gamma (VBF)} = r_{VBF} b_{\gamma \gamma},\quad \mu_{\gamma \gamma (ggF)} = r_{ggF} b_{\gamma \gamma}.
\end{equation}
In principle these can both be measured separately, as discussed in \sref{expstatus}, in which case a large discernible difference compared to the SM should be found if EWBG takes place within the MSSM.  We also consider searches sensitive to both ggF and VBF, with possibly different efficiencies $\xi$ for each production mode. In this case case the $\gamma \gamma$ signal strength prediction (similarly for other channels) is given by
\begin{equation}
\label{e.muinclusive}
\mu_{\gamma \gamma} = \frac{r_{ggF} + r_{VBF} r_{SM}}{1+r_{SM}} b_{\gamma \gamma} \ \ ,  \  \ \ \ \mathrm{where} \ \ \ \ \ \ r_{SM} = \frac{\xi_\mathrm{VBF}}{\xi_\mathrm{ggF}} \cdot   \frac{ \ \sigma_{SM}(VBF)}{\ \sigma_{SM}(gg \rightarrow h)},
\end{equation}
and $r_{SM} \sim 0.1$ if the efficiencies for ggF and VBF are comparable.

While EWBG fixes the stops of the MSSM to particular values, in principle the value of $m_A$ can range from the decoupling limit to very low values of $m_A$ as in Bino-driven EWBG\cite{BinoDrivenEWBG}, while still preserving the successes of EWBG.  The value of $m_A$ is important since it alters the $VVh$ coupling compared to the SM, which in turn rescales
\begin{equation}
r_{VBF}\ , \  r_{AP} \ \approx \  \sin(\beta - \alpha_{eff}),
\end{equation}
where $\alpha_{eff}$ is the effective $CP$-even Higgs mixing angle.  (In the decoupling limit, $\alpha_{eff} = \beta - \pi/2$ thus, $r_{VBF}, r_{AP} \approx 1$ in this case.) Our analysis will take this allowed range for $m_A$ into account.  For technical details on the decay width and cross section ratio computation, as well as the associated theoretical errors, the reader is referred to the Appendix.

\begin{figure}
\begin{center}
\begin{tabular}{lcc}
\begin{sideways}$\ \ \ \ \  \ \  m_A = 2 \tev$\end{sideways}
& &
\includegraphics[width=14cm]{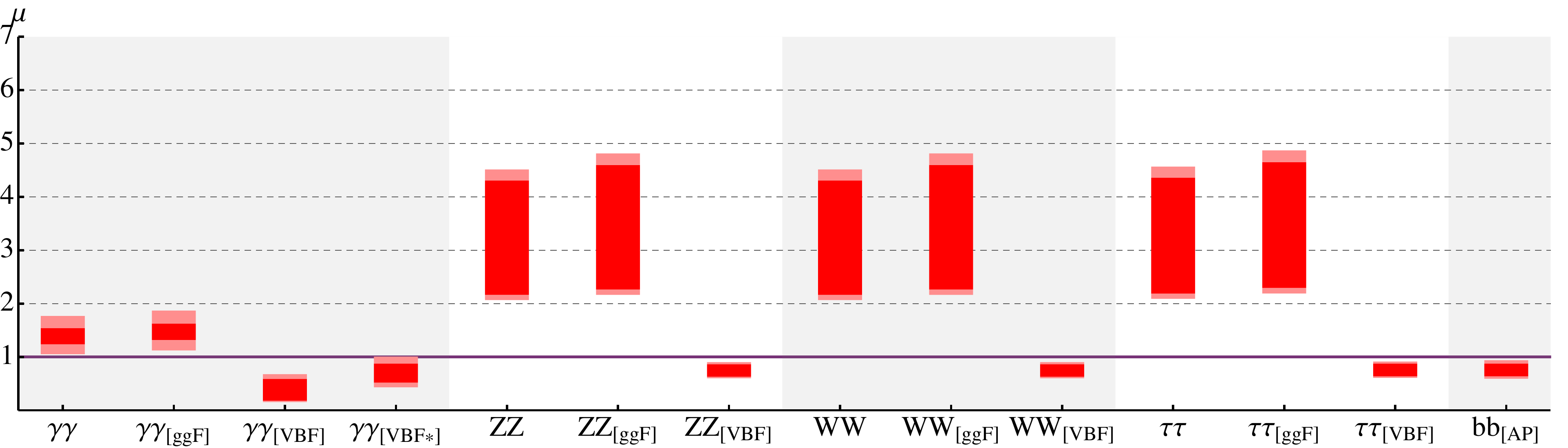}
\\ \\
\begin{sideways}$\ \ \ \ \ \   \  m_A = 300 \gev$\end{sideways}
&  &
\includegraphics[width=14cm]{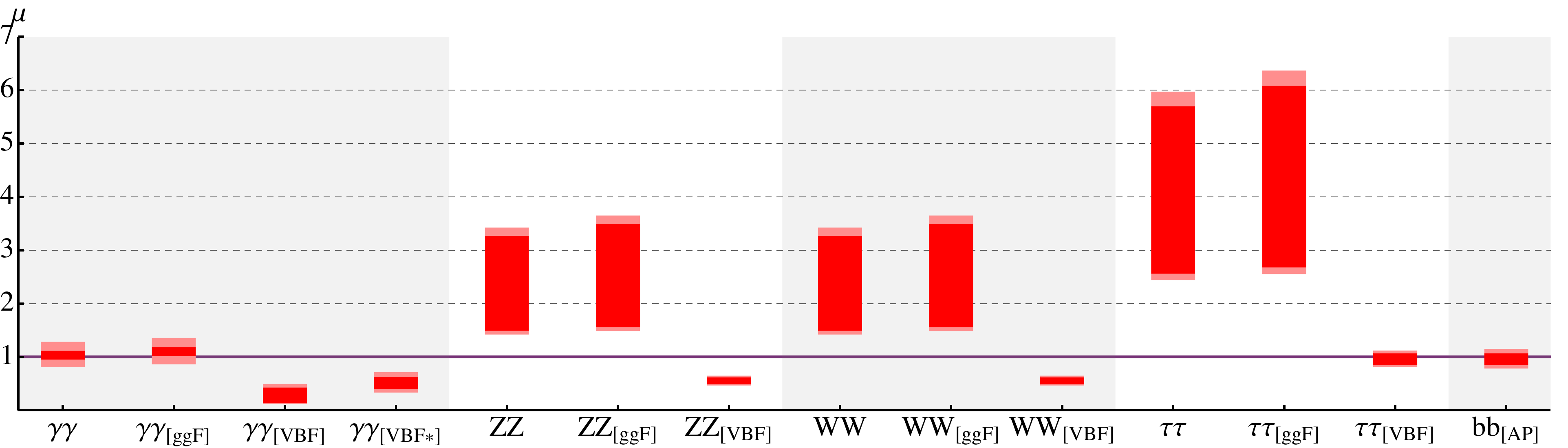}
\end{tabular}
\end{center}
\caption{Theoretical EWBG fingerprint for $m_A=2 \tev$ and 300 GeV, for a range of stop masses from 80-115 GeV including theory errors. Shown are signal strength predictions for each channel, with subscripts indicating an exclusive production mode. The exception is $\gamma \gamma_{[\mathrm{VBF}*]}$, which denotes the signal strength prediction for a $ h \rightarrow \gamma \gamma$ search with VBF cuts \cite{CMSgammagamma}, such that $\xi_\mathrm{VBF}/\xi_\mathrm{ggF} \approx 30$ in \eref{muinclusive}. The purple line is the SM expectation. This fingerprint is for $m_h = 125 \gev$, but the dependence on $m_h$ is very small in the $123 - 128 \gev$ neighborhood of Higgs masses.  Solid red bands indicate the range of predictions for $m_{\tilde t_R} \in (80, 115) \gev$. The light red bands indicate the theory error at the top and bottom of the stop mass range. $\tan \beta$ was allowed to vary in the range $(5, 15)$, but its effect is very small since $m_h$ was taken as a low-energy input. The rate of decays that are dominated by gluon fusion increases for lighter stop masses, while $\gamma \gamma$ and channels sensitive to Vector Boson Fusion and Associated Production are much less affected. 
}
\label{f.fingerprint}
\end{figure}

Combining the shifts for $b_{\gamma\gamma}$ and $r_{ggF}$ from  (\ref{e.hggdecay}) and (\ref{e.hgagadecay}) with all the other channels allows us to  map the theoretical  fingerprint of the entire EWBG scenario over the allowed range of stop masses.  We show this fingerprint in \fref{fingerprint} for the decoupling limit and for $m_A = 300 \gev$. The decoupling limit is required for generic $CP$-violating phases, and the latter value of $m_A$ was chosen because, as we will see in the next section, it \emph {minimizes} the tension between experimental data and the EWBG prediction. What is striking about \fref{fingerprint} is that even including a range of stop masses, theory uncertainties, and $m_A$, there are easily discernible correlations amongst the various channels.  

While not all of these channels have been measured to a very precise level, the particular fingerprint of EWBG in the MSSM means that this scenario can in principle be ruled out by combining information from the 7 TeV LHC Higgs searches only. Of course, it is possible that the Higgs may not ultimately be found to have a mass of $\approx 125 \gev$. However, because the deviations from SM are so large for EWBG, it is still possible to bound EWBG for arbitrary Higgs mass given the current data sets. 

We will examine this in detail in the following sections, but the fingerprint detailed in \fref{fingerprint} stresses the power of setting limits even with channels that may not have enough data for a discovery with the entire 2012 8 TeV data set. Eventually, the study of correlations such as these could be one of our most powerful tools into discerning the effects of new physics, if it is related to electroweak symmetry breaking.


\section{Experimental Status}
\label{s.expstatus} \setcounter{equation}{0} \setcounter{footnote}{0}

We will start by briefly outlining the available experimental data before moving on to show the extent to which different regions of EWBG parameter space are excluded, both with and without the assumption of a $\approx 125 \gev$ Higgs.

\subsection{Available Data}
\label{ss.availabledata}

\begin{table}
\begin{center}
\hspace*{-6mm}
\begin{tabular}{cc|c|c|c|c||c|c|}
\cline{3-8}
& & \multicolumn{4}{|c||}{Production Mode Sensitivity} & \multicolumn{2}{|c|}{Signal Strength Bounds} \\
\cline {3-8}
& & ggF & VBF & AP & Inclusive & Source & $m_h$ range (GeV)\\
\hline  \hline
\multicolumn{1}{|c|}{\multirow{2}{*}{$\gamma \gamma$}}
&
\multicolumn{1}{|c|}{ATLAS \cite{ATLASgammagamma}} &  &  &  & $\star$  & official & (110, 150)\\
\cline{2-8}
\multicolumn{1}{|c|}{}
&
\multicolumn{1}{|c|}{CMS \cite{CMSgammagamma, CMSmoriond}} & $\star$ & $\star$ &  & & reconstructed$^\dagger$ \cite{approxLogL1} & (120, 128)    \\
\hline \hline
\multicolumn{1}{|c|}{\multirow{2}{*}{$Z Z^*$}}
&
\multicolumn{1}{|c|}{ATLAS \cite{ATLASZZ}} &  &  &  & $\star$ & official & (110, 150)\\
\cline{2-8}
\multicolumn{1}{|c|}{}
&
\multicolumn{1}{|c|}{CMS \cite{CMSZZ}} &  &  &  & $\star$  & reconstructed$^\dagger$ \cite{approxLogL1} & (120, 128)\\
\hline \hline
\multicolumn{1}{|c|}{\multirow{2}{*}{$W W^*$}}
&
\multicolumn{1}{|c|}{ATLAS \cite{ATLASWW}} &  &  &  & $\star$   & official & (110, 150)\\
\cline{2-8}
\multicolumn{1}{|c|}{}
&
\multicolumn{1}{|c|}{CMS} & $\circ$  & $\circ$  &  &  & --- & ---    \\
\hline \hline
\multicolumn{1}{|c|}{\multirow{3}{*}{$b b$}}
&
\multicolumn{1}{|c|}{ATLAS \cite{ATLASbb}} &  &  &  $\star$ & &  official & (110, 130)\\
\cline{2-8}
\multicolumn{1}{|c|}{}
&
\multicolumn{1}{|c|}{CMS \cite{CMSbb, CMSmoriond}} &  &  &  $\star$ &  & reconstructed$^\dagger$ \cite{approxLogL2}    & (110, 130)\\
\cline{2-8}
\multicolumn{1}{|c|}{}
&
\multicolumn{1}{|c|}{D0 + CDF \cite{Tevatronbb}} &  &  &  $\star$ & &  official & (100, 150)     \\
\hline \hline
\multicolumn{1}{|c|}{\multirow{2}{*}{$\tau \tau$}}
&
\multicolumn{1}{|c|}{ATLAS \cite{ATLAStautau}} &  & $\circ$ &   & $\star$ & reconstructed \cite{approxLogL2} & (110, 150)\\
\cline{2-8}
\multicolumn{1}{|c|}{}
&
\multicolumn{1}{|c|}{CMS \cite{CMStautau}} & $\circ$ & $\circ$ &  &   & --- & (110, 150) \\
\cline{1-8}
\end{tabular}
\caption{Summary of the relevant higgs searches and their sensitivity to production modes and decay channels.  '$\star$' indicates that the experiment released sufficient experimental data for our analysis. '$\circ$' indicates that even though search channel was considered in the experiment, the publicly available data was insufficient for our analysis.  Whenever official signal strength bounds were unavailable we performed our analysis using approximate reconstructed likelihoods for the signal strength (which are likely to give more conservative bounds than the official fit). For CMS $\gamma \gamma$, $ZZ^*$ we used likelihoods supplied to us by the authors of \cite{approxLogL1}, while for CMS $bb$ and both $\tau \tau$ searches we reconstructed the likelihoods using the methods of \cite{approxLogL2}.  $^\dagger$CMS made official $\gamma \gamma_{[\mathrm{VBF}]}$, $ZZ^*$, $bb_{[\mathrm{AP}]}$ signal strength bounds available at $m_h = 124, 125$ \gev, which were used instead of the reconstructed approximations.
}
\label{t.channels}
\end{center}
\end{table}

 \begin{figure}[h]
\begin{center}
\hspace*{-15mm}
\begin{tabular}{lcc}
& $m_h = 125 \gev$  & $m_h = 126 \gev$\\
\begin{sideways}$\ \ \ \  \ \ m_A = 2 \tev$\end{sideways}
&
\includegraphics[width=8.5cm]{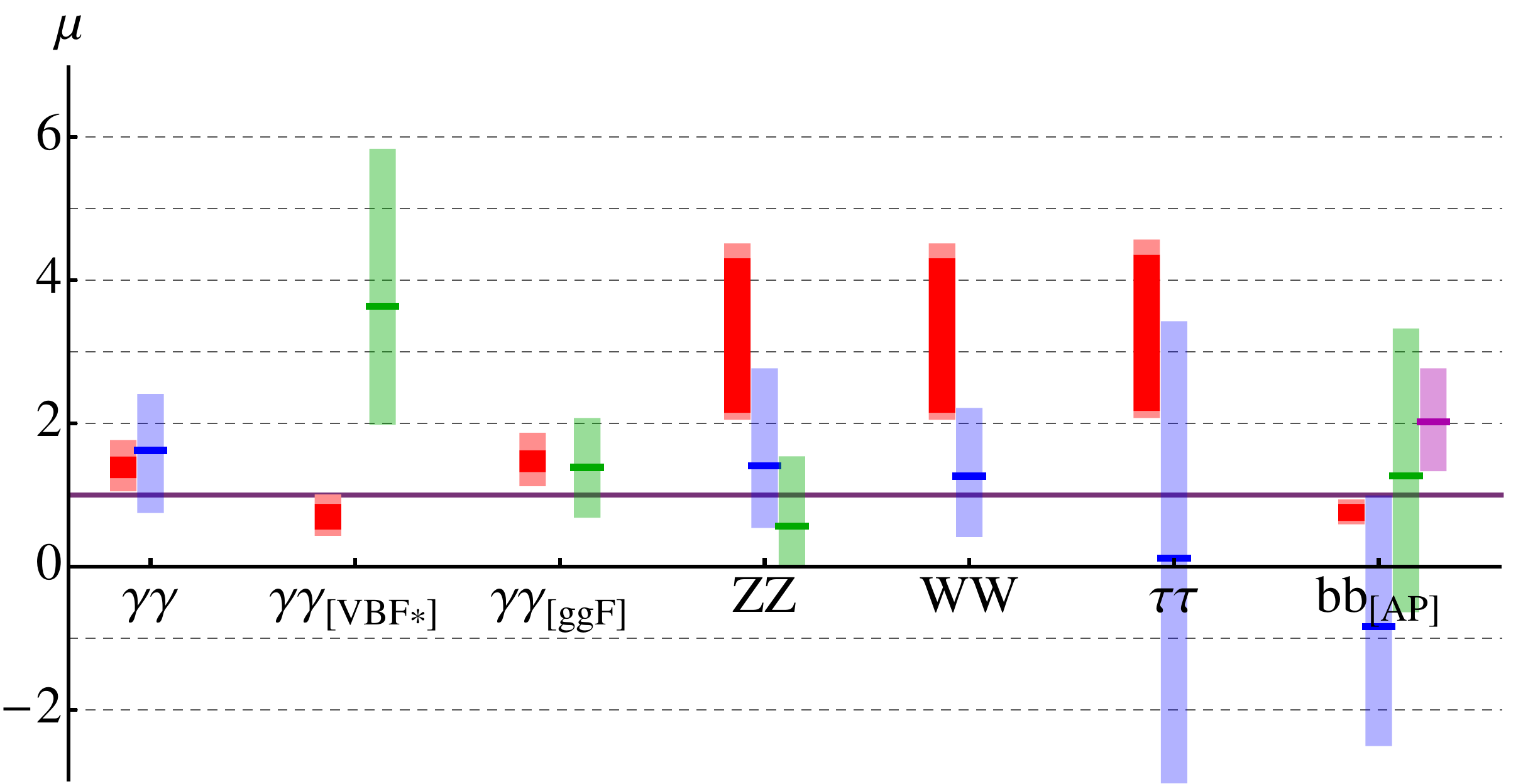}
& 
\includegraphics[width=8.5cm]{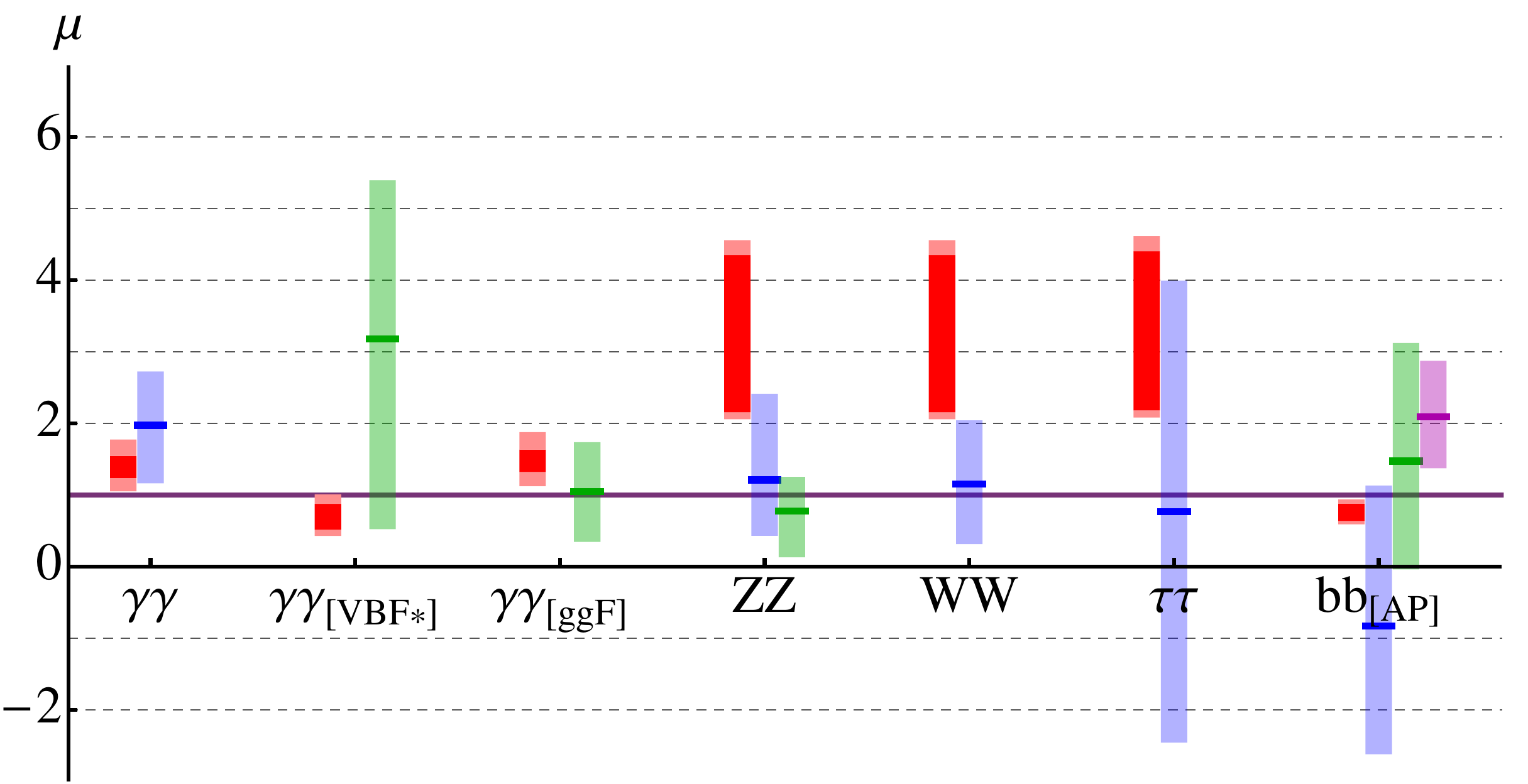}
\\ \\ 
\begin{sideways}$ \ \ \ \ \  m_A =  300 \gev$\end{sideways}
& \includegraphics[width=8.5cm]{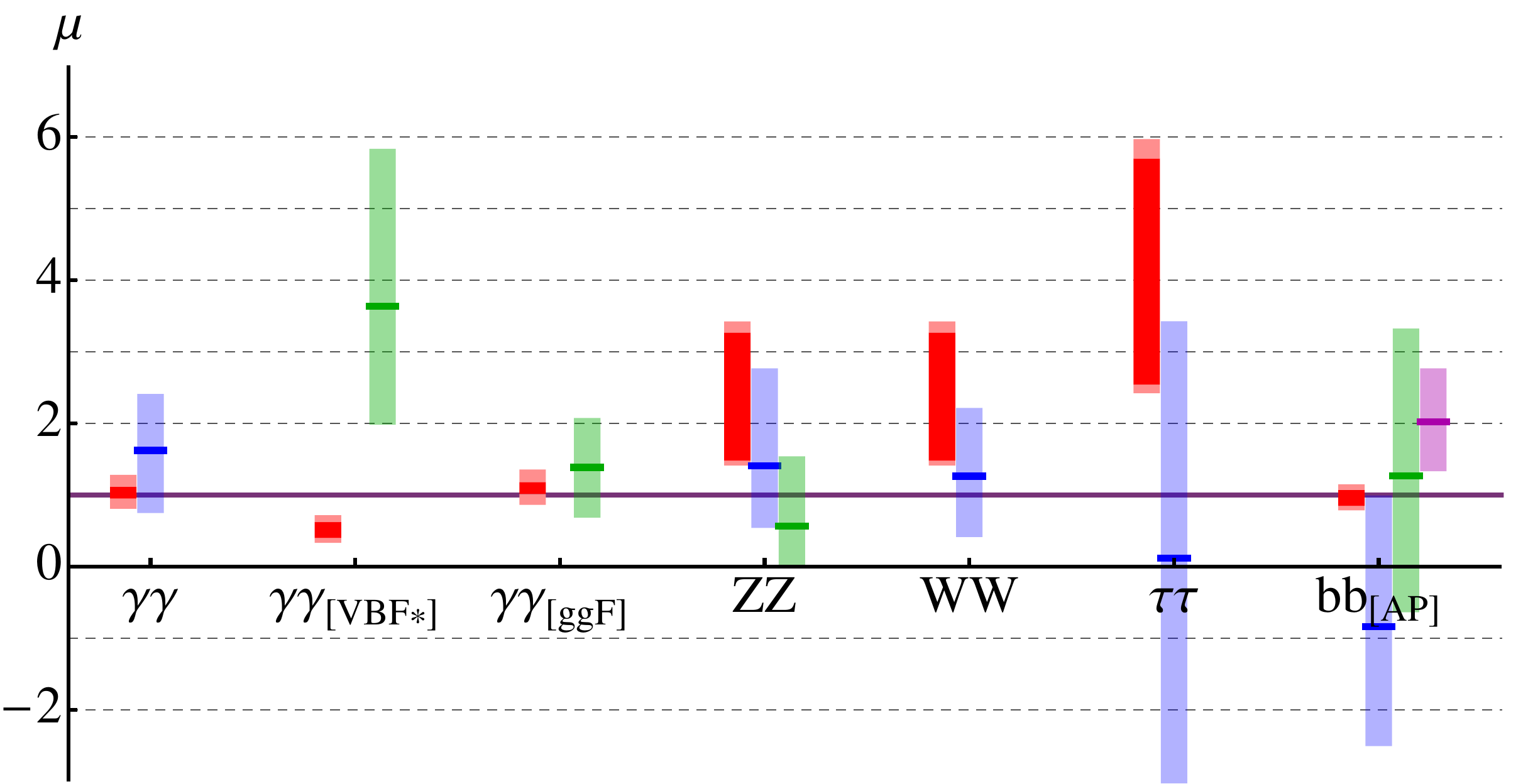}
& 
\includegraphics[width=8.5cm]{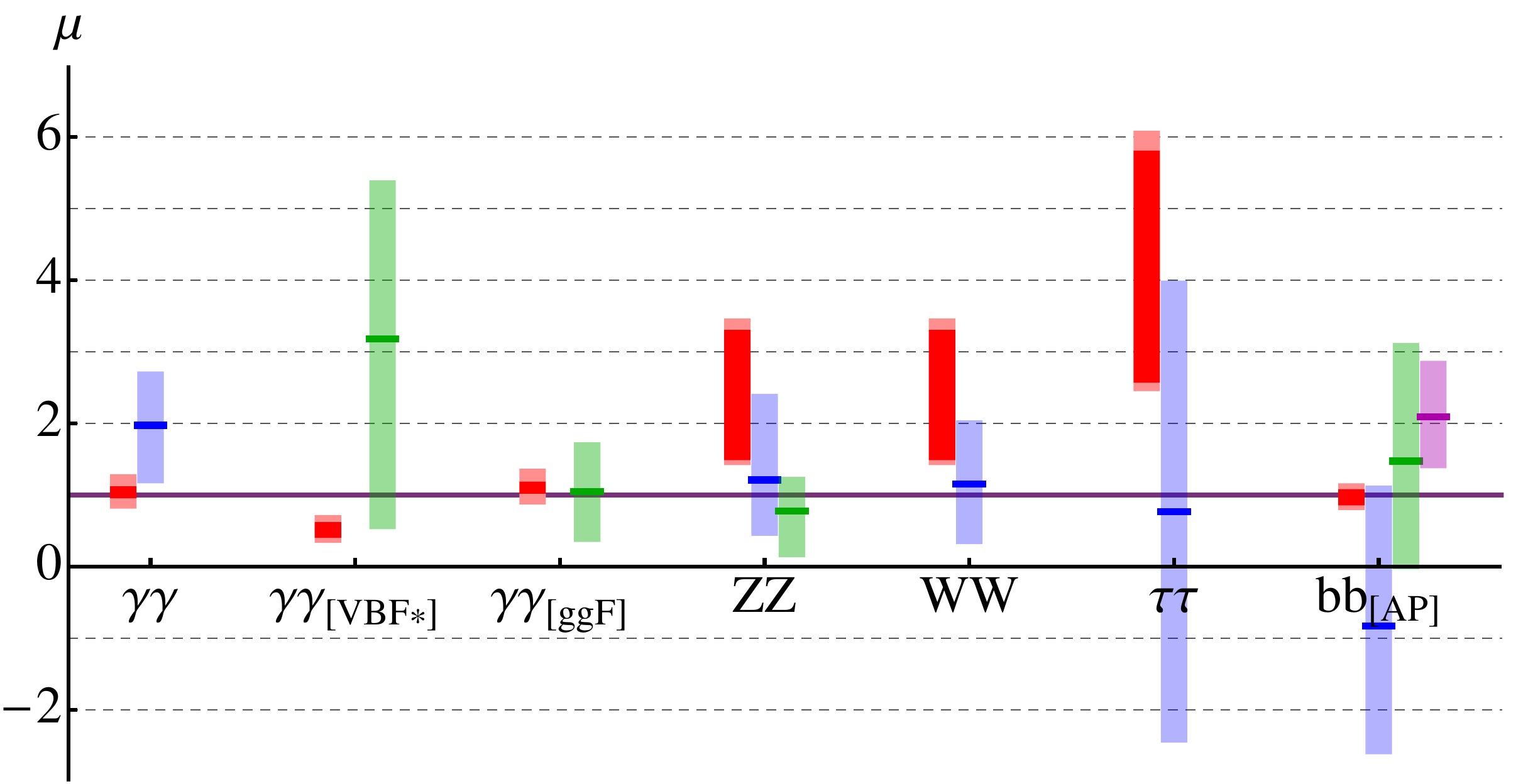}
\end{tabular}
\end{center}
\caption{
Comparing the signal strength predictions  for each Higgs decay channel of electroweak baryogenesis in the MSSM with the ATLAS, CMS and Tevatron data as explained in \ssref{availabledata}. Subscripts indicate exclusive production via a single mode. For each channel we show up to three bands:  the EWBG prediction, with $m_{\tilde t_R} \in (80, 115) \gev$ (red, with theoretical error bands in light red) and the ATLAS/CMS 1-$\sigma$ best-fit measurements (blue/green, with central value indicated in dark blue/green). In the $bb_{[\mathrm{AP}]}$ channel we also show the combined Tevatron constraint as a fourth band (purple). The SM prediction is indicated with a horizontal line at $\mu = 1$. }
\label{f.resultplot}
\end{figure}

\tref{channels} summarizes all the available Higgs searches to date that are relevant to our analysis. A few remarks are in order:
\begin{itemize}

\item  We use Eqns. (\ref{e.partialsigstrength2}) and (\ref{e.muinclusive}) to compute the theory predictions for the $\gamma \gamma$ signal strengths (similarly for the other channels). The inclusive signal strength prediction assumes equal signal efficiencies for ggF and VBF, which is a conservative choice for setting limits. Production is dominated by ggF, but  assuming the VBF efficiency to be zero would lead us to slightly overestimate the theory prediction for the signal strength $\mu$, since ggF is enhanced in our MSSM scenario. As we will see, this  would increase tension with the data. Therefore, we set the two efficiencies to be equal, while noting that some deviation from this assumption will not invalidate the analysis since $r_{SM}$ is small for this search. For the $ h \rightarrow \gamma \gamma$ search with VBF cuts we set $\xi_\mathrm{VBF}/\xi_\mathrm{ggF} = 30$ in \eref{muinclusive} \cite{CMSgammagamma}.

\item Official signal strength bounds were not always available for each channel. Fortunately, the authors of \cite{approxLogL1} reconstructed approximate signal strength likelihoods for the CMS $\gamma \gamma$, $Z Z^*$ searches by using the information that is publicly available and generating their own event samples.  For other searches we used the methods of \cite{approxLogL2}, very similar to the ideas of \cite{approxLogL1}, to reconstruct approximate likelihoods where necessary.

\item We used the older ATLAS  $h \rightarrow WW^* \rightarrow \ell \ell \nu \nu$ search using $2.05 \  \mathrm{fb}^{-1}$ of data \cite{ATLASWW} rather than the updated version with $4.7 \  \mathrm{fb}^{-1}$ \cite{ATLASWWnew}. The latter is significantly more constraining and looks to increase the tension with the EWBG prediction, but there is not enough information available to reliably disentangle the ggF and VBF contributions. The CMS $WW^*$ search \cite{CMSWW} is omitted because signal strength bound are only reconstructible for $m_h = 120, 130 \gev$.

\end{itemize}

\fref{resultplot} compares the signal strength predictions to the experimental signal strength bounds in all available channels, for $m_h = 125$ and $126 \gev$. The results are displayed for these two higgs masses since they are preferred by the CMS and ATLAS $\gamma \gamma$ searches, respectively. A visual inspection already reveals some tension with the data. We will now make this impression quantitative.

\subsection{Excluding Electroweak Baryogenesis in the MSSM}
\label{ss.excludingEWBG}
Given the large error bars in the early Higgs data it is perhaps surprising that we can make relatively strong statements regarding the exclusion of electroweak baryogenesis. This is due to the correlations of the signal strength predictions in the various channels and their dependence on EWBG parameters.

The Higgs signal in the various channels depends only very weakly on $\tan \beta$, since our parameterization takes $m_h$ as a low-energy input and $\tan \beta$ can not be large for successful EWBG. Therefore, for a given Higgs mass, the parameter space of EWBG in the MSSM is the $(m_A, m_{\tilde t_R})$ plane. Once the Higgs mass is determined this will be the relevant parameter space to exclude.

After taking into account theory error and the small amount of $\tan \beta$ dependence, each point in the  $(m_A, m_{\tilde t_R})$ plane maps to a range of signal strength vectors $(\mu_{\gamma \gamma}, \mu_{\gamma \gamma_\mathrm{VBF*}}, \ldots)$, which constitute the range of experimental predictions for this parameter point. Maximizing the signal strength likelihood function  $L(\mu_{\gamma \gamma}, \mu_{\gamma \gamma_\mathrm{VBF*}},  \ldots)$, which is obtained from experimental data, over the range of allowed signal strength vectors gives the exclusion for this point in the $(m_A, m_{\tilde t_R})$ plane. Unfortunately the signal strength likelihood function $L$ is not directly available. However, we can obtain a passable approximation by first assuming that the separate searches are independent, and then using the 1-$\sigma$ best-fit bounds on the separate signal strengths to obtain gaussian approximations for $L_i(\mu_i)$ (taking into account asymmetric error bars where appropriate). Normalizing $\log L = \sum_i L_i(\mu_i)$ to have a maximum value of zero, we obtain the desired likelihood function.

\begin{figure}
\begin{center}
\includegraphics[width=9cm]{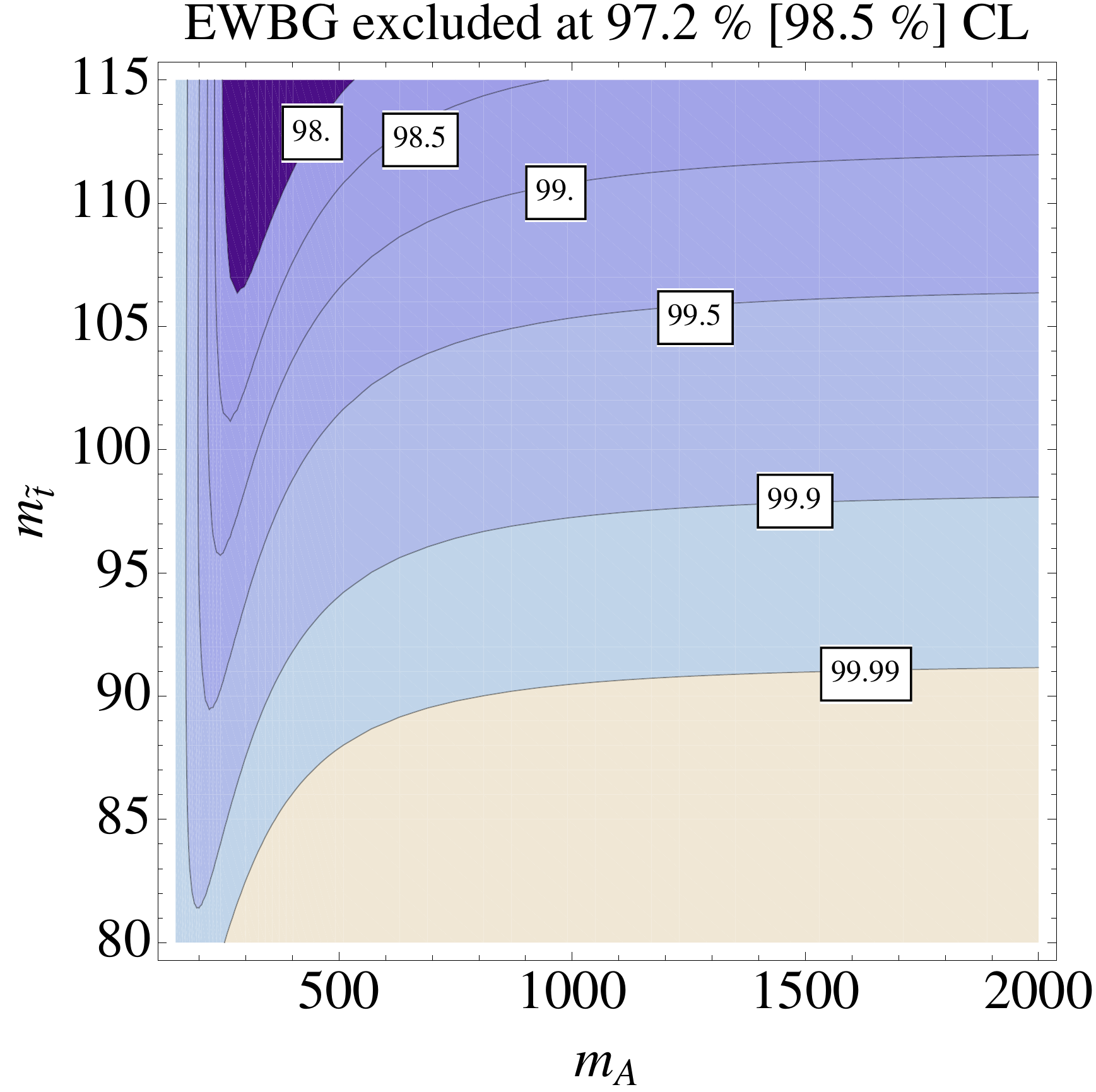}
\vspace*{-4mm}
\end{center}
\caption{
Exclusion plot of EWBG parameter space for $m_h = 125 \gev$, obtained by combining the signal strength bounds from the various ATLAS and CMS Higgs searches  (not Tevatron) as outlined in \ssref{availabledata}.  The smallest exclusion at $m_A \approx 300 \gev$, $m_{\tilde t_R} = 115 \gev$ is 97.2\%, which increases to 98.5\% if we enforce the decoupling limit ($m_A > 1 \tev$).
}
\label{f.exclusionplot}
\end{figure}

In \fref{exclusionplot} we show the exclusion across EWBG parameter space, obtained by combining ATLAS and CMS data for $m_h = 125 \gev$. The entire parameter space is excluded at the 97.2 \% CL (98.5 \% if we enforce the decoupling limit). The least excluded points are at $m_A \approx 300 \gev$, with $m_{\tilde t_R}$ as high as possible. One could consider several variations on this plot, each with similar results.
\begin{itemize}
\item Until the Higgs mass is precisely known one should generally consider the exclusion of EWBG over a range of $m_h$. This produces a very similar pattern of exclusion to \fref{exclusionplot}, with tension minimized for $m_A \approx 300 \gev$ and high $m_{\tilde t_R}$. The exclusion as a function of Higgs mass (from ATLAS and CMS data only) is:
\begin{center}
\begin{tabular}{l|cccccc}
$m_h$ in GeV & 123 & 124 & 125 & 126 & 127 & 128\\
\hline
minimal exclusion ($\%$) for all $m_A$: & 93.5 & 95.9   &  97.2 & 92.1  &  92.6  &  90.2
\\
minimal exclusion ($\%$) for $m_A > 1 \tev$: & 99.5 & 98.0 & 98.6 &  98.7 &  99.9 &  99.99
\end{tabular}
\end{center}
We can see that  EWBG is always excluded with a CL beyond 90 \%.

\item There appears to be a slight mismatch between the most favored Higgs mass from the CMS and ATLAS $\gamma \gamma$ searches.   If we were to shift the $\gamma\gamma$ channel signal strength bounds for ATLAS or CMS by 1 GeV, the exclusions would go up to 98\% for $m_h=125,126$ GeV.

\item One could ask how these exclusions would change if we also made use of the Tevatron $bb$ constraint. Not surprisingly, it does not change significantly:
\begin{center}
\begin{tabular}{l|cccccc}
$m_h$ in GeV & 123 & 124 & 125 & 126 & 127 & 128\\
\hline
minimal exclusion ($\%$) for all $m_A$: & 90.8 & 95.5   &  97.2  & 93.5 &  94.1  &  92.4
\\
minimal exclusion ($\%$) for $m_A > 1 \tev$: & 99.6 & 98.5   & 99.0  & 99.3 & 99.97  & 99.999
\end{tabular}
\end{center}

\item  It is instructive to consider the exclusion obtained by combining \emph{only} the two $\gamma \gamma$ constraints, each at their respective best-fit Higgs masses. $m_A < 500 \gev$ is significantly disfavored, since the reduced $\gamma \gamma h$ effective coupling exacerbates the tension between the  $\gamma \gamma [\mathrm{VBF}]$ signal strength prediction (already lower than SM) and the larger-than-SM observation by CMS. Over the entire EWBG parameter space, the exclusion from only $\gamma \gamma$ data is 89.5\%. 
\end{itemize}

EWBG is more excluded for smaller stop masses because  lighter stops lead to greater enhancement of the Higgs production cross-section. This increases the $\gamma \gamma$, $ZZ^*$ and $WW^*$ signal strengths, causing tension with the observations. For large $m_A$, the signal strength predicted by EWBG is somewhat larger that the observed value for the channels $Z Z^*$ and $W W^*$, which leads to relatively strong exclusion. As we reduce $m_A$, the Higgs couplings to $\gamma \gamma$, $Z Z$ and $W W$ decrease. The reduced signal strength leads to weaker exclusion from $ZZ$ and $WW$, but  as explained above the increasing tension in the $\gamma \gamma [\mathrm{VBF}]$ channel strongly disfavors very small $m_A$. This leads to the `sweet spot' of $m_A$ around 200 - 300 GeV.

If we believe that recent excesses observed in the various LHC and Tevatron searches are due to a Higgs in the 123 - 128 GeV mass range, then these early measurements already exclude EWBG in the MSSM at the 90\% CL. If we combine only the $\gamma \gamma$ observations at the best-fit higgs mass values, the exclusion is 89.5\%. While more data is needed for a  definite conclusion, it is clear that EWBG in the MSSM is  strongly disfavored even at this early stage.

\subsection{Excluding a more general Light-Stop Scenario}
\label{ss.generalLSS}

\begin{figure}
\begin{center}
\hspace*{-10mm}
\begin{tabular}{ccc}
\includegraphics[width=8cm]{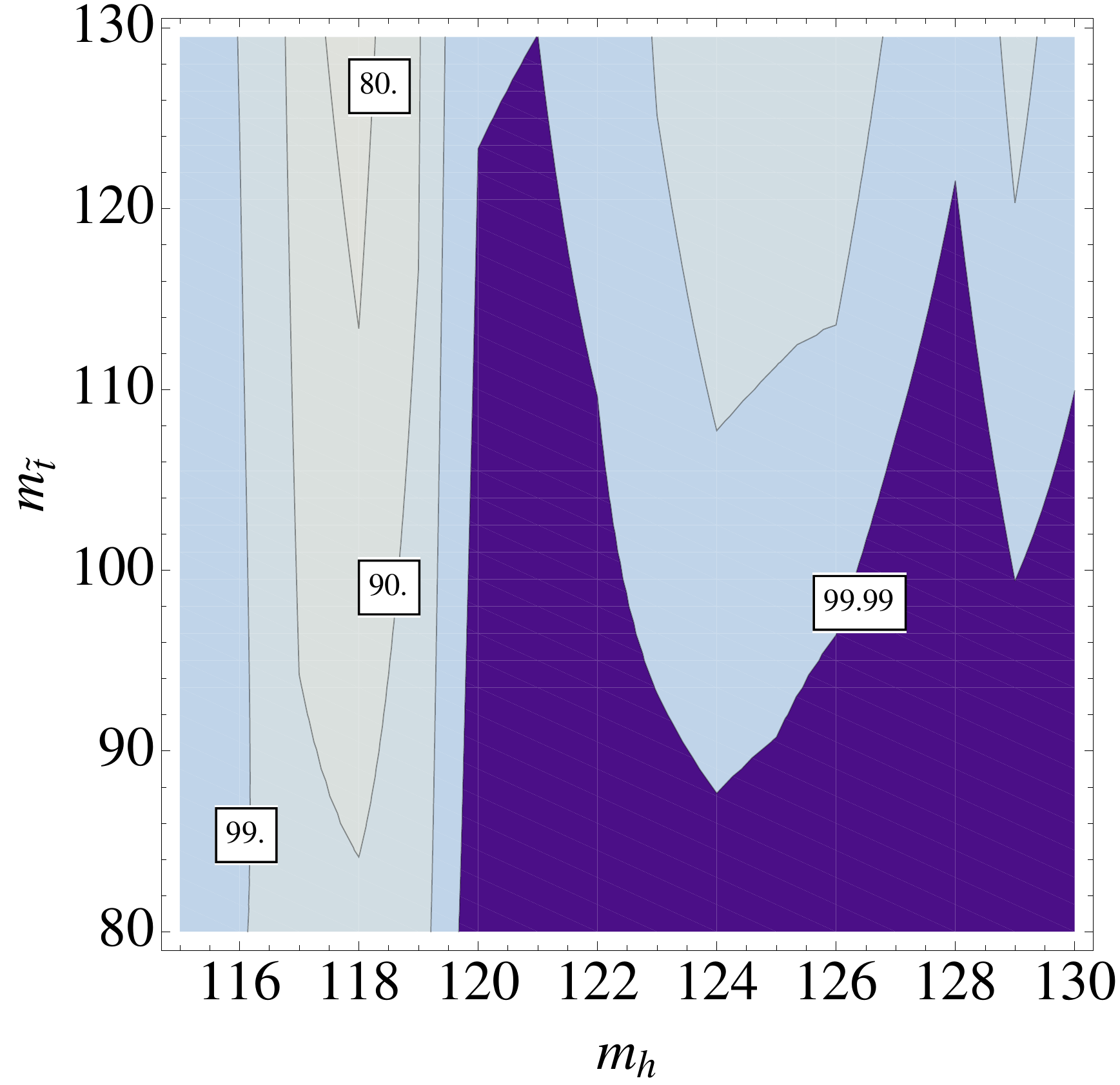}
&
&
\includegraphics[width=8cm]{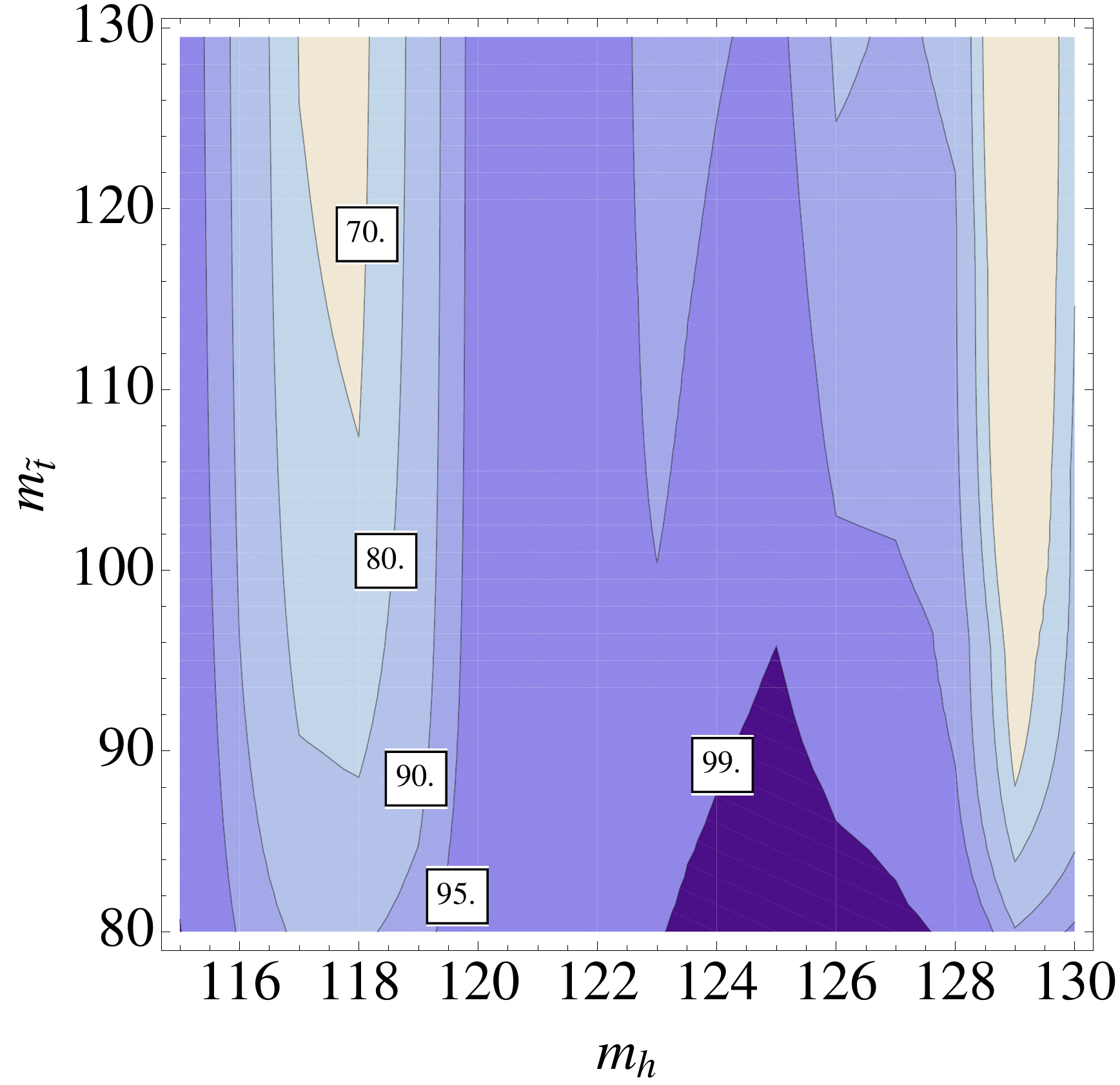}
\\
(a) Decoupling Limit & & (b) $150 \gev < m_A < 2 \tev$
\end{tabular}
\end{center}
\caption{Exclusion of a more general Light Stop Scenario in the $(m_h, m_{\tilde t_R})$ plane. As before, $\tilde t_L$ is taken to be very heavy, while $m_A$ and $\tan \beta$ were varied in the range $(150, 2000) \gev$ and $(5, 15)$. This exclusion plot was created via the same method as \fref{exclusionplot}, using both ATLAS and CMS data but not the Tevatron $bb$ bound. For each point in the $(m_h, m_{\tilde t_R})$ plane we minimize exclusion with respect to theory error, $\tan \beta$ dependence and $m_A$ dependence. The decoupling limit $m_A > 1 \tev$ is enforced in (a), while (b) allows the whole range of $m_A$.
}
\label{f.mhmstopexclusion}
\end{figure}

\label{ss.generalLSSexclusion}
One could  loosen the assumptions of our analysis, and ask what the available LHC data tells us about a wider range of Higgs and stop masses. Dropping the assumption of a 123 - 128 GeV Higgs allows us to examine the prospects of electroweak baryogenesis in the MSSM if the Higgs were to sit at a different mass. 

\fref{mhmstopexclusion} shows the exclusion from ATLAS and CMS data as a function of the $(m_h, m_{\tilde t_R})$ plane. This exclusion plot was created via the same method as \fref{exclusionplot}, using gaussian approximations of the signal strength bounds. For each point in the $(m_h, m_{\tilde t_R})$ plane we minimize exclusion with respect to theory error, $\tan \beta$ dependence and $m_A$ dependence, using  the experimental signal strength bounds 
for whatever Higgs masses they are available (see \tref{channels}). However, there is one additional complication with this expanded Higgs mass range: the ATLAS $ZZ$ bounds have extremely asymmetric error bars for $m_h < 122 \gev$. This suggests a reduced reliability of the gaussian likelihood approximation, and therefore we do not use the ATLAS $ZZ$ bounds for $m_h < 122 \gev$.

What does  \fref{mhmstopexclusion} imply for MSSM EWBG in general? Without a Higgs mass constraint, the successful electroweak phase transition requires $m_{\tilde t_R} \lsim 120 \gev$ and $m_h < 128 \gev$ \cite{CNQWfullstudy}. As we can see, LHC data already excludes almost all of this parameter space at more than $90\%$ CL, with the most notable exception being a (117 - 119) GeV Higgs mass window  that is only excluded at $70 - 85 \%$ CL. 

Including Tevatron data does not significantly enhance these constraints. If we were to include the ATLAS $ZZ$ constraints below $m_h = 122 \gev$ by naively applying its $+1\sigma$ error bar to $-1\sigma$ and using the gaussian likelihood approximation, then the exclusion of the allowed Higgs Mass window increases by $\approx 10\%$. However, given the possibly long tail of the likelihood distribution these numbers must be taken with a grain of salt. 

At any rate, it is clear that the LHC Higgs searches are already excluding large portions of the LSS parameter space, even without the assumption of a particular Higgs mass. Further data should allow exclusion of the 117 - 119 GeV Higgs mass window, but these considerations are of course irrelevant if a Higgs mass of $\approx 125 \gev$ is confirmed. We also point out that our exclusions could likely be improved if the experimental collaborations provided more detailed signal strength likelihoods. 

Finally, we point out that the discovery or exclusion of the  MSSM pseudoscalar Higgs would of course significantly enhance the power of these constraints by restricting the allowed value of $m_A$.

\section{Conclusion}
\label{s.conclusion} \setcounter{equation}{0} \setcounter{footnote}{0}

EWBG in the MSSM has long been an attractive possibility for baryogenesis due to the presence of additional light supersymmetric states within reach of colliders.  With the 7 TeV run of the LHC now completed, there finally is enough high energy data to confront the entire parameter space of EWBG.  However, in direct searches there are in principle ways to weaken constraints without affecting the physics of EWBG.  In this paper we looked at the indirect consequences for the Higgs sector alone, which represents an irreducible constraint on EWBG in the MSSM\footnote{This is irreducible in the sense that to alter the effects of indirect constraints would require introducing new couplings to the Higgs directly and thus by definition alter the predictions for EWBG in the MSSM.}.  Given that the only window for EWBG that still exists is the LSS, there are very large corrections to Higgs phenomenology and in particular strong correlations amongst channels as we have demonstrated.  

We have shown that in the context of a possible Higgs mass measurement of  $m_h \approx 125 \gev$ at the LHC,  EWBG in the MSSM is now ruled out at greater than a 98\% CL in the decoupling limit for the Higgs sector, and at least 90\% CL for lighter values of $m_A$. 
This is primarily due to the fact that this heavy of Higgs mass even further restricts the allowed parameter space for EWBG in the MSSM.  If we relax the constraint of $m_h \approx 125 \gev$, to allow for a larger parameter space, we can {\em still} constrain the LSS in full generality with the current LHC data.  We find that the only significant region excluded at less than $90 \%$ CL  is $m_h \approx 117$ - $119 \gev$. This window can be excluded with more data, though of course the point is moot if the Higgs is confirmed to have a mass of $\approx 125 \gev$. 

Finally, one can speculate that LHC Higgs searches could in fact offer an almost model-independent way to exclude electroweak baryogenesis, even beyond the MSSM implementation.\footnote{For a recent example, see \cite{CohenPierce}.} At its very basic level, the mechanism of EWBG requires the presence of weak-scale particles with substantial Higgs couplings to generate the strong first-order phase transition via their contributions to the thermal potential. These particles, via those very same couplings, could then significantly contribute to the effective Higgs couplings and allows for testing EWBG in other scenarios~\cite{futurework}. While complete exclusion of the electroweak baryogenesis idea may not be possible \cite{nonexcludableEWBG}, this could nonetheless reduce the array of viable models to a few very special cases. Ultimately, the indirect tests of Higgs phenomenology using correlations, such as those we have demonstrated in this paper, may prove to be the most important window into new physics.

\subsection*{Acknowledgements}

We would like to thank Sally Dawson for many helpful conversations. We are also very grateful to Tomer Volansky and Eric Kuflik for supplying us with their reconstructed CMS signal strength likelihoods from \cite{approxLogL1}, as well as useful discussions.  The work of D.C. was supported in part by the National Science Foundation under Grant PHY-0969739. The work of P.J. was supported in part by the U.S.  National Science Foundation under grant NSF-PHY-0969510, the LHC Theory Initiative, Jonathan Bagger, PI. The work of P.M. was supported in part by NSF CAREER Award NSF-PHY-1056833.

 \appendix

\section*{Appendix: Higgs Decay Rate Calculations}
\label{a.appendix} \setcounter{equation}{0} \setcounter{footnote}{0}

Most of the public programs available for branching ratio and Higgs production/decay calculations of the MSSM Higgs bosons are unsuitable for the unusual stop spectrum under consideration. We have therefore implemented a simple \texttt{Mathematica} code to perform the computations following the references \cite{Djouadi:2005gi, Djouadi:2005gj}. This is feasible since we only care about calculating decay widths and production cross sections from a very simple low-energy spectrum, not the derivation of that spectrum from UV-parameters.

The case of small $m_A$ is handled as follows:
For each choice of $m_h, m_{\tilde t_R}, \tan \beta$ and $m_A$, we use the expressions in \cite{Djouadi:2005gj} to derive the LH stop mass required to give the desired Higgs mass at one-loop resummed order. This allows us to compute the radiatively corrected charged and heavy neutral Higgs masses at the same order and include their contributions in the light Higgs decay widths and branching fractions. This one-loop resummed calculation might be insufficient if we were actually interested in how exactly the Higgs spectrum is derived from a low-energy theory, but for the purposes of `sweeping through the Higgs spectrum' as we change $m_A$ this is certainly expected to work well.

 \subsubsection*{Decay Widths \& Branching Fractions}

The branching ratio calculations in our code include most of the important higher order corrections under the assumption of no stop mixing. Understanding the theoretical uncertainties is quite important for a comparison with LHC data. Therefore we briefly summarize the decay width calculation for each channel and estimate the theory error from those higher order corrections we neglected, as well as chargino and neutralino contributions. 

\begin{itemize}

\item $h \rightarrow g g$ : The LO contribution is a one-loop diagram with quarks or squarks running in the loop. Only the light right-handed stops were considered in our calculation since the contribution from the heavy scalars decouples. Most of the important NLO/NNLO QCD corrections \cite{Chetyrkin:1997iv, Spira:1995rr, Dawson:1993qf} were implemented  with the exception of those from gluino loops. However, the latter are quite small (a few percent for $M_g \sim 1$ TeV) \cite{Harlander:2003bb}. The electroweak (EW) corrections \cite{Djouadi:1994ge, Degrassi:2004mx} are $ \lsim 5 \% $   due to large cancellations from various contributions.  

\item $h \rightarrow \gamma \gamma/\gamma Z$ : This channel is also loop induced at LO. However, in addition to quarks and squarks, there is also a small contribution from charginos and charged Higgs in the loop. While the contributions are small for $m_{\chi^+} \gsim 250$ GeV, they can be as big as $10 \%$ for $m_{\chi^+} \sim 100$ GeV \cite{Djouadi:1996pb, hgaga-SUSY}. The contributions from the charged Higgs are ignored since they are assumed to be heavy and therefore decouple. Most of the NLO/NNLO QCD corrections  \cite{Spira:1995rr, hgaga-QCD-NLO} are implemented, an exception being the QCD corrections to the squark loop diagrams leading to an error of $\sim 10 \%$ \cite{Djouadi:1996pb}. The EW corrections from the top Yukawa coupling \cite{hgaga-EW-NLO} is also implemented for the $h \rightarrow \gamma \gamma$ channel.

\item $h \rightarrow V V^*$ : The vector boson $V$ here refers to the massive gauge bosons $W$ and $Z$. Since, we assume that the Higgs mass is $m_h \approx 125$ GeV, one of the final state vector bosons is always off-shell \cite{hVV-LO}. The NLO EW corrections \cite{EW-NLO, hVV-EW-NLO} have been implemented for this channel and we expect the remaining theoretical uncertainties to be small. In the decoupling limit the $hVV$ couplings are SM-like,  so the partial decay width is identical to the SM case. For small $m_A$ the tree-level coupling is rescaled (see below). The NLO SUSY contributions have been ignored as they are expected to be small.     
  
\item $h \rightarrow f \bar{f}$ : For a $125$ GeV Higgs, the decay to top quarks is kinematically forbidden and thus the relevant channels are decays to bottom, charm, and tau. We have implemented NLO/NNLO QCD corrections \cite{hff-QCD-NLO} including the running of the quark masses to absorb the large logarithms \cite{hff-runningmass}. NLO EW contributions have also been implemented \cite{EW-NLO, hff-EW-NLO}. In the MSSM, the bottom quark Yukawa coupling gets significant corrections which are non-vanishing in the decoupling limit \cite{hff-SUSY-QCD}. In our scenario, the corrections to the bottom quark Yukawa are given by $\Delta_b \propto \alpha_s M_g \mu \tan\beta / M_{\tilde{b}_L}^2$. Since the left-handed sbottom is extremely heavy  we conclude that these SUSY QCD corrections are small and can be ignored. The only significant source of error in this channel is the uncertainty in the quark masses since they can not be directly measured. This translates to an error of $\sim 4$ $\%$ for the channel $h \rightarrow b \bar{b}$  \cite{Dittmaier:2012vm}.
\end{itemize}

\subsubsection*{Production Cross Section Ratios}
The LO cross-section for a $2 \rightarrow 1$ process is proportional to the decay width of the inverse process. Therefore we can use the following approximation to estimate the MSSM Higgs production cross-section through gluon fusion:
\begin{equation*}
r_{ggF} \equiv \frac{\sigma_{MSSM}(gg \rightarrow h) }{\sigma_{SM}(gg \rightarrow h)} \approx \frac{\Gamma_{MSSM} (h \rightarrow g g)}{\Gamma_{SM} (h \rightarrow g g)} 
\end{equation*}
where we use the decay widths calculated above. The QCD K-factors for $ h \rightarrow g g$ differ by $\sim 6$ $\%$ for the SM and the MSSM case and this difference is taken into account by using NLO decay widths. Thus we expect this approximation to work very well, and take its contribution to the theory error of our signal strength prediction to be small (compared to the other sources of uncertainty), of order a few percent. 

The Vector Boson Fusion and Associated Production cross section ratios are $r_{VBF}\ , \  r_{AP} \approx 1$ in the decoupling limit, but for small  $m_A$ the  $VVh$ tree-level couplings are rescaled compared to the SM. Therefore
\begin{equation*}
r_{VBF}\ , \  r_{AP} \ \approx \  \sin(\beta - \alpha_{eff}),
\end{equation*}
where $\alpha_{eff}$ is the effective $CP$-even Higgs mixing angle.  (In the decoupling limit, $\alpha_{eff} = \beta - \pi/2$ thus, $r_{VBF}, r_{AP} \approx 1$ in this case.) We expect the error introduced by this approximation to also be a few percent.  

Finally, to compute theory predictions for inclusive signal strengths we need
\begin{equation*}
r_{SM} = \frac{\sigma_{SM}(VBF)}{\sigma_{SM}(gg \rightarrow h)} \sim 0.1.
\end{equation*}
The SM Higgs production cross section and associated theoretical errors are calculated in  \cite{higgsxsection}, which we use to obtain predictions and theoretical uncertainties for $r_{SM}$.

\end{document}